# Morphognosis: the shape of knowledge in space and time

Thomas E. Portegys

tom.portegys@ey.com Ernst & Young LLP, New York, NY, USA

## Abstract
Artificial intelligence research to a great degree focuses on the brain and behaviors that the brain generates. But the brain, an extremely complex structure resulting from millions of years of evolution, can be viewed as a solution to problems posed by an environment existing in space and time. The environment generates signals that produce sensory events within an organism. Building an internal spatial and temporal model of the environment allows an organism to navigate and manipulate the environment. Higher intelligence might be the ability to process information coming from a larger extent of space-time. In keeping with nature's penchant for extending rather than replacing, the purpose of the mammalian neocortex might then be to record events from distant reaches of space and time and render them, as though yet near and present, to the older, deeper brain whose instinctual roles have changed little over eons. Here this notion is embodied in a model called *morphognosis* (*morpho* = shape and *gnosis* = knowledge). Its basic structure is a pyramid of event recordings called a *morphognostic*. At the apex of the pyramid are the most recent and nearby events. Receding from the apex are less recent and possibly more distant events. A morphognostic can thus be viewed as a structure of progressively larger chunks of space-time knowledge. A set of morphognostics forms long-term memories that are learned by exposure to the environment. A cellular automaton is used as the platform to investigate the morphognosis model, using a simulated organism that learns to forage in its world for food, build a nest, and play the game of Pong.

## Introduction
The human brain is the seat of intelligence. Thus when we attempt to craft intelligence, naturally we turn to it as a guide. Fortunately, neuroscience is proceeding at an astounding pace (Kaiser, 2014; Stetka, 2016), methodically unpacking its mysteries. Yet the complexity of the brain, with billions of neurons and trillions of synapses, remains daunting. Teasing apart which aspects and features of the brain are essential to the function of intelligence and which are incidental is a crucial and difficult task. Unfortunately, the prospects of understanding complex systems through examination and dissection are questionable (Jonas and Kording, 2016). And as for constructing a complete precise brain model, it is possible, as John von Neumann believed (Mühlenbein, 2009), that at a certain level of complexity the simplest precise description of a thing is the thing itself. In reaction to this, some efforts, such as The Human Brain Project (2015) and Numenta (Hawkins, 2004; White paper, 2011), have taken the position that analysis must be complemented with synthesis and simulation to achieve a satisfactory level of understanding.

From an artificial intelligence (AI) viewpoint, we must keep in mind that the purpose of a brain is to allow an organism to navigate and manipulate its environment. Thus it is a solution to problems posed by the environment. The approach of this project is to model the environment as something that could plausibly be in turn modeled by an artificial brain.

Some researchers maintain that the environment largely consists of a body for the brain to interact with. The embodied brain will thus leverage the sensory and motor capabilities of a body that are adapted to an environment. Robotics researchers such as Brooks (1999), Hoffmann and Pfeifer (2011) have argued that true artificial intelligence can only be achieved by machines that have sensory and motor skills and are connected to the world through a body. However, this approach belies the problem since the body, like the brain, is also a solution to its environment.

Determining a model of an organism's environment is more tractable than creating a brain model of an environmental model. But it requires settling on what is in the world that produces sensory events and reacts to motor responses. Confounding this is that we of course must use our brains to do this. There is a common and somewhat ironic tendency to describe AI inputs and outputs in human cognitive terms, i.e. post-processed brain output, such as symbolic variables.

Hoffman (2009) argues that evolution has shaped our senses and perceptual machinery to only provide information on events that are ancestrally significant, such as finding food and safety. Other events in the environment that we cannot directly sense must be mapped through technology onto our sensory capabilities. For example, in the age of science the existence and use of X-rays is important, but we sense them only indirectly, as shadows on photographic film. Indeed, Hoffman argues that reality may be more radically alien than we can imagine.

Epistemological offerings would seem at best too abstract to be useful for framing a sensory-response environment, and at worst useless, as in the cases of nihilism and solipsism. And physics has in recent times become increasingly muddier on the "true" nature of reality:

- The arrow of time may be related to the perception of entropy (Halliwell, 2011).
- String theory demands a number of extra infinitesimal dimensions (Rickles, 2014).
- The perception of space may be a holographic projection (Bousso, 2002).
- Reality could be a cellular automaton (Wolfram, 2002), a graph (Wolfram, 2015), or a simulation (Moskowitz, 2016).

Despite these hazards, people universally experience the environment as a space-time structure. And even if there is a different underlying substructure, the model is empirically effective. The presence of mammalian brain structures for mapping spatial events (Vorhees and Williams, 2014) provides evidence for the processing of this type of information. Similarly, brain structures for sensing the passage of time have also found support (Sanders, 2015).

Using space-time as a model, it can be speculated that higher intelligence is the ability to process information arising from a larger extent of space-time. And in keeping with nature's penchant for extending rather than replacing, the purpose of the mammalian neocortex might then be to record events from distant reaches of space and time and render them, as though yet near and present, to the older, deeper brain whose instinctual roles have changed little over eons. If this is so, these structures would be repurposed to embody language and abstract concepts.

Building an internal spatial and temporal model of the environment allows an organism to navigate and manipulate the environment. This paper introduces a model called *morphognosis* (morpho = shape and gnosis = knowledge). Its basic structure is a pyramid of event recordings called a *morphognostic*, as shown in Figure 1. At the apex of the pyramid are the most recent and nearby events. Receding from the apex are less recent and possibly more distant events.

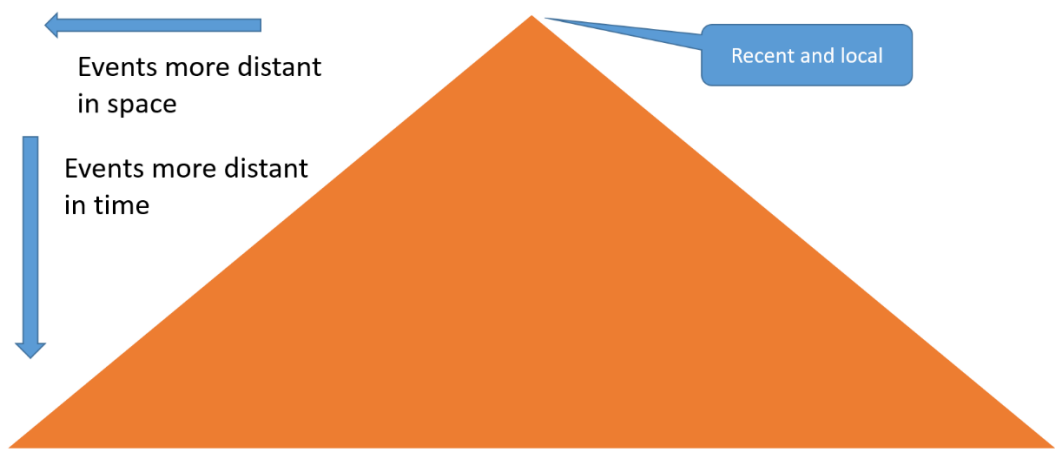

Figure 1 – Morphognostic event pyramid

Morphognosis is partially inspired by an abstract morphogenesis model called *morphozoic* (Portegys et al., 2017). Morphogenesis is the process of generating complex structures from simpler ones within an environment. Morphozoic is based on hierarchically nested neighborhoods within a cellular automaton. Morphozoic was found to be robust and noise tolerant in reproducing a number of morphogenesis-like phenomena, including Turing diffusion-reaction systems (Turing, 1952), gastrulation, and neuron pathfinding. It is also capable of image reconstruction tasks.

## Description

The morphognosis model is demonstrated in three 2D cellular environments: (1) a food foraging task, (2) a nest building task, and (3) the game of Pong. The food foraging task is used as a venue to further define the model.

### Food foraging

In this task a virtual creature called a *mox* finds itself in a 2D cellular world as shown in Figure 2. To find food the mox must navigate around various obstacles of various types (colors).

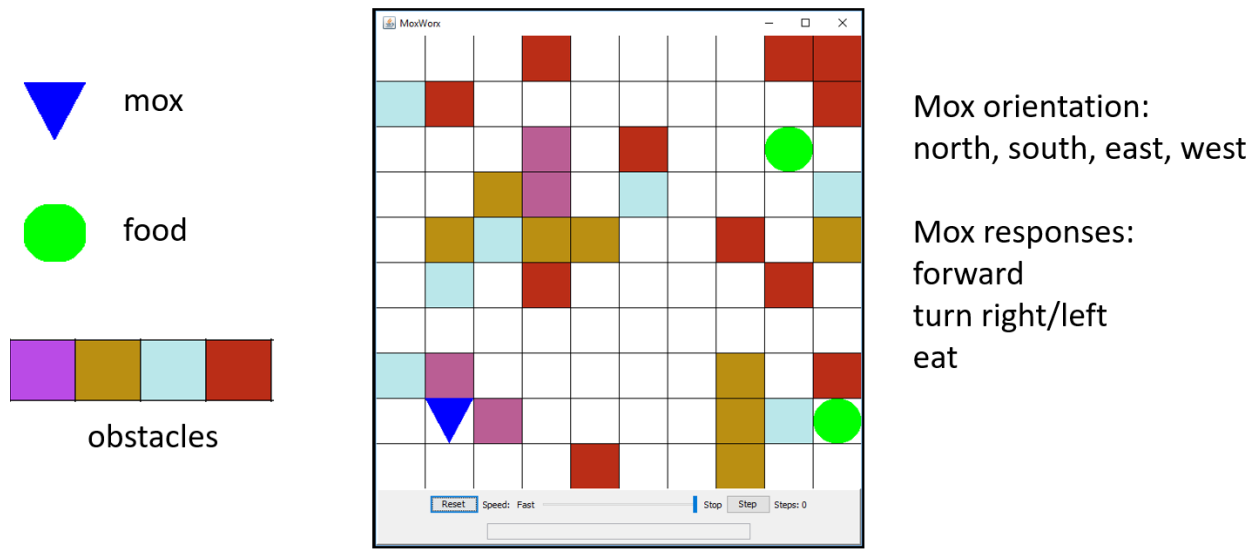

Figure 2 - Mox food foraging in a 2D cellular world.

Figure 3 shows a snapshot of a morphognostic describing the space-time events, in this case obstacle encounters, while the mox forages. In this case a neighborhood is configured as a 3x3 set of sectors.

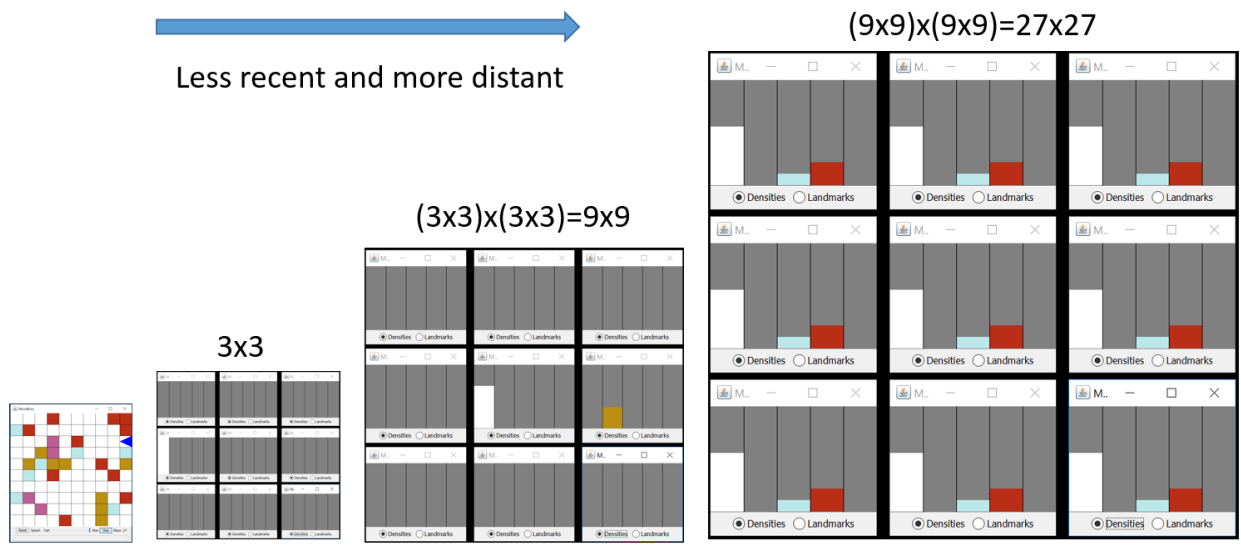

Figure 3 - Pyramid of obstacle type densities arranged as hierarchy of 3x3 cell neighborhoods.

Cell type densities are stored instead of raw cell values to allow linear scaling of information as the hierarchy increases since storing individual cell values would result in a geometric growth. The cell type density is only one of a number of possible statistical or aggregation functions that could be used. An alternative might be to look at the distribution of cell types as an image processing operation, such as taking a Laplacian, Sobel or other image operator.

## Morphognostic spatial neighborhoods

A cell defines an elementary neighborhood:

$neighborhood_0 = cell$

A non-elementary neighborhood consists of an *NxN* set of *sectors* surrounding a lower level neighborhood:

$neighborhood_i = NxN(neighborhood_{i-1})$

where *N* is an odd positive number.

The value of a sector is a vector representing a histogram of the cell type densities contained within it:

$value(sector) = (density(cell\text{-}type_0), density(cell\text{-}type_1), ... density(cell\text{-}type_n))$

The number of cells contributing to the density histogram of a sector of $neighborhood_i = N^{i-1}xN^{i-1}$

## Morphognostic temporal neighborhoods

A neighborhood contains events that occur between time *epoch* and *epoch + duration*:

$t1_0 = 0$
$t2_0 = 1$
$t1_i = t2_{i-1}$
$t2_i = (t2_{i-1} * 3) + 1$
$epoch_i = t1_i$
$duration_i = t2_i - t1_i$

## Metamorphs

In order to navigate and manipulate the environment, it is necessary for an agent to be able to respond to the environment. A *metamorph* embodies a morphognostic→response rule. A set of metamorphs can be learned from a manual or programmed sequence of responses within a world.

Metamorphs establish an important feedback:

- Learned morphognostics shape responses.
- Responses shape the learning of morphognostics.

Metamorph "execution" consists generating a morphognostic for the current mox position and orientation then finding the closest morphognostic contained in the learned metamorph set, where:

$distance(metamorph_i, metamorph_j) =$

$$\sum_{x}^{neighborhoods} \sum_{y}^{sectors} \sum_{z}^{cell\ types} abs(cell\ type\ density_{i,x,y,z} - cell\ type\ density_{j,x,y,z})$$

*Artificial neural network implementation*

In a complex environment, generating a large number of metamorphs may be prohibitive in terms of storage and search processing. Alternatively, metamorphs can be used to train an artificial neural network (ANN), as shown in Figure 4, to learn responses associated with morphognostic inputs. During operation, a current morphognostic can be input to the ANN to produce a learned response. The ANN also has these advantages:

- Faster.
- More compact.
- More noise tolerant.

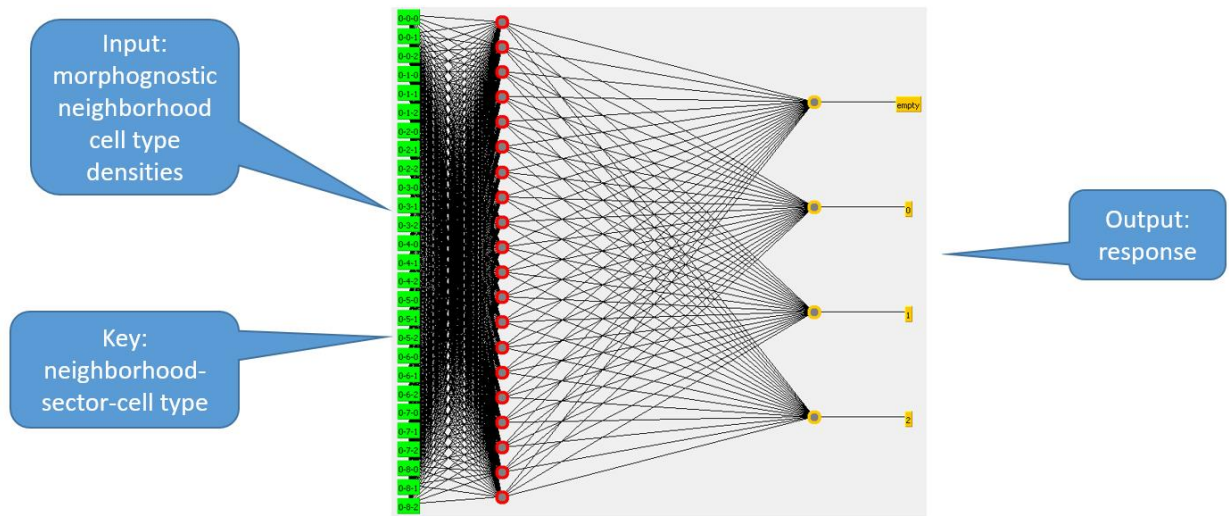

Figure 4 – Metamorph artificial neural network.

## Results

The mox were trained in worlds featuring a number of randomly placed obstacles of various types. Training was done by "autopiloting" the mox along an optimal path to the food. This generated a set of metamorphs suitable for testing. Table 1 shows the results of varying the neighborhood hierarchy depth in a 10x10 world. Success indicates the mean amount of food eaten, so 1 is a perfect score. It can be observed that more obstacles tend to improve performance. This is because they tend to form unique landmark configurations to guide the mox. Larger neighborhoods also tend to improve performance.

| Neighborhoods | Obstacle types | Obstacles | Food |
|---|---|---|---|
| 1 | 1 | 10 | 0.1 |
| 1 | 1 | 20 | 0.2 |
| 1 | 2 | 10 | 0 |
| 1 | 2 | 20 | 0 |
| 1 | 4 | 10 | 0 |
| 1 | 4 | 20 | 0 |
| 2 | 1 | 10 | 0.3 |
| 2 | 1 | 20 | 0.4 |
| 2 | 2 | 10 | 0.2 |
| 2 | 2 | 20 | 0.6 |
| 2 | 4 | 10 | 0.2 |
| 2 | 4 | 20 | 0.6 |
| 3 | 1 | 10 | 1 |
| 3 | 1 | 20 | 0.9 |
| 3 | 2 | 10 | 1 |
| 3 | 2 | 20 | 1 |
| 3 | 4 | 10 | 1 |
| 3 | 4 | 20 | 1 |

Table 1 – Foraging in a 10x10 world.

The next test examines how well the model performs when the test world is not a duplicate of a training world, but is similar to a set of training worlds. Thus for this, multiple training runs are used. Before each training run, the cell types of all the cells are probabilistically modified to a random value. A successful test run must then rely on a composite of multiple training runs. The results are shown in Table 2. Of note is how performance only begins to falter under heavy noise and few training runs.

| Noise | #Train | Food |
|---|---|---|
| 0.1 | 1 | 1 |
| 0.1 | 5 | 1 |
| 0.1 | 10 | 1 |
| 0.25 | 1 | 0.9 |
| 0.25 | 5 | 1 |
| 0.25 | 10 | 1 |
| 0.5 | 1 | 0.6 |
| 0.5 | 5 | 0.8 |
| 0.5 | 10 | 0.9 |

Table 2 – Foraging with noise.

## Nest building

This task illustrates how the morphognosis model can be used to not only navigate but also manipulate the environment. Figure 5 left shows an environment in which a nest will be constructed out of 4 stones (reddish circles) on top of an elevation depicted by the shaded cells. The mox must seek out the stones, pick them up, and assemble them into the completed nest shown in Figure 5 right.

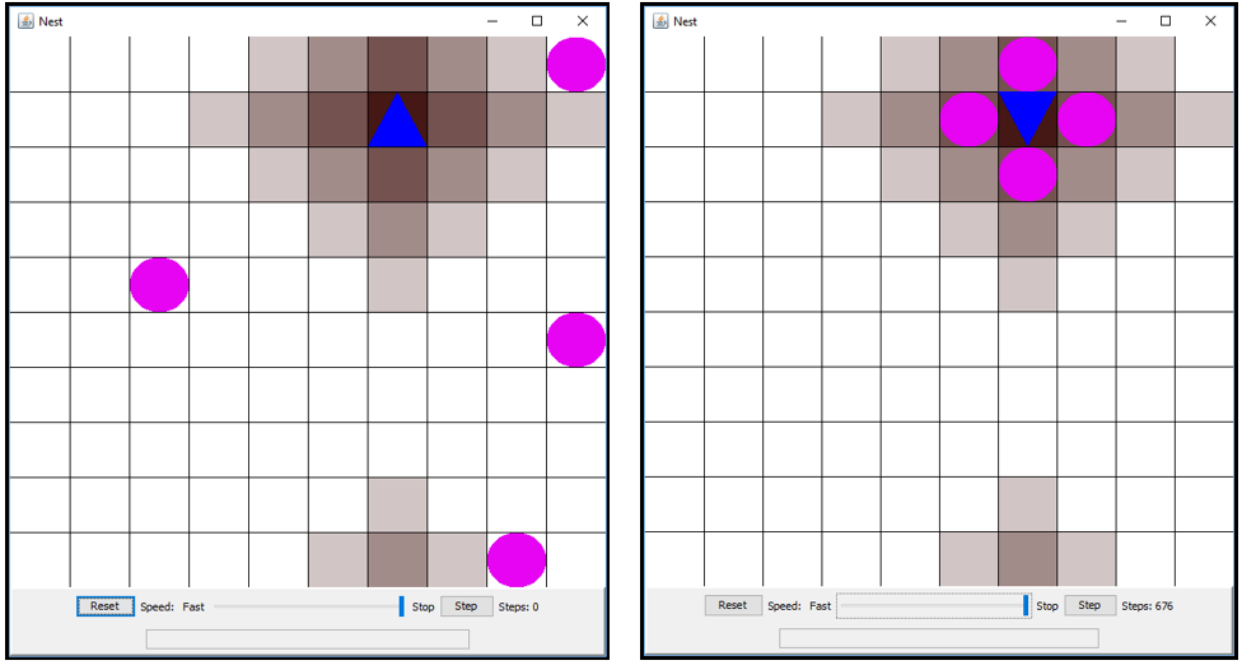

Figure 5 – Nest building with gathered stones. Left: scattered stones. Right: completed nest.

For this task, the mox is capable of sensing the presence of a stone immediately in front of it, and sensing the elevation gradient both laterally and in the forward-backward direction. In addition to the forward and turning movements used by the foraging task, the mox is capable of picking up a stone in front of it and dropping the stone onto an unoccupied cell in front of it. It also senses whether it is carrying a stone.

Training was done by running the mox through 10 repetitions on "autopilot" to build a set of metamorphs. The environment was then reset and the mox tested to discover whether it is capable of building the nest. Over 50 trials were performed with 100% success. Internally, the sensory information from the stone, gradient and stone carry states were sufficient to achieve success with a neighborhood hierarchy of only one level.

## Pong game

Much of the real world is nondeterministic, taking the form of unpredictable or probabilistic events that must be acted upon. If AIs are to engage such phenomena, then they must be able to learn how to deal with nondeterminism. In this task the game of Pong poses a nondeterministic environment. The learner is given an incomplete view of the game state and underlying deterministic physics, resulting in a nondeterministic game.

## Game details
- The goal of the game is to vertically move a paddle to prevent a bouncing ball from striking the right wall, as shown in Figure 6.
- Ball and paddle move in a cellular grid.
    - Unseen deterministic physics moves ball in grid.

- Cell state: (ball state, paddle state)
    - Ball state: (empty, present, moving left/right/up/down)
    - Paddle state: (true | false)
- Learner orientation: (north, south, east, west)
- Responses: (wait, forward, turn right/left)
    - If paddle present and orientation north or south, then forward response moves paddle also.

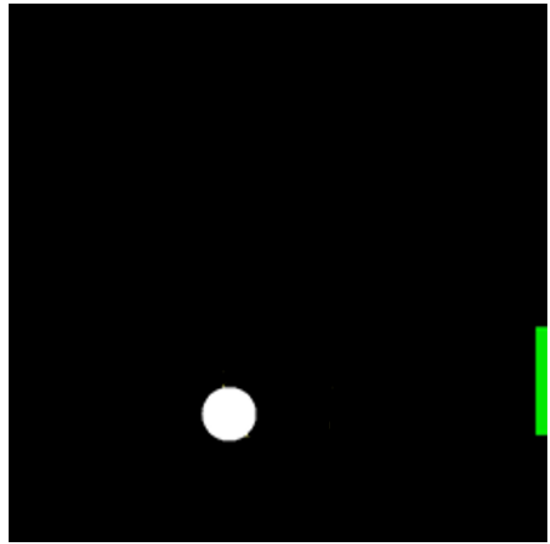

Figure 6 – The game of Pong.

## Procedure and results
- Learner was trained with multiple randomly generated initial ball velocities.
- When the ball moved left and right, the learner moved with the ball.
- When the ball moved up or down, the learner moved to the paddle and moved it up or down.
    - This was the challenge: remembering ball state while traversing empty cells to the paddle so as to move it correctly, then to turn and return to ball for next input.

Testing on random games: 100% successful.

## Conclusion
This is an early exploration of the morphognosis model. The positive results on the three tasks prompt future investigation. Possible next tasks include:

- Web building. Can a space-time memories of building one or more training webs allow one to be built in a quasi-novel environment?
- Food foraging social signaling. Bees retain memories of foraging food sources that they communicate to other bees through instinctive dancing. Can this task be cast into the model?

The Java code is available at https://github.com/portegys/MoxWorx